\documentclass[floatfix,
 aps,
 prb,
 reprint,
 amsmath,amssymb,
 superscriptaddress
]{revtex4-2}

\usepackage{tikz}
\usetikzlibrary{positioning, arrows.meta, calc, fit, fadings}
\tikzfading[name=densfade, inner color=transparent!0, outer color=transparent!100]
\usepackage{amsmath}
\usepackage{bm}
\usepackage{hyperref}
\usepackage{booktabs}
\usepackage{siunitx}
\usepackage{newtxtext,newtxmath} 

\newcommand{\Exc}{E_{\mathrm{xc}}}
\newcommand{\vxc}{v_{\mathrm{xc}}}
\newcommand{\br}{\textbf{r}}

\begin{document}

\title{\textbf{Machine-learned Laplacian-level density functional from exact exchange-correlation potentials and energies} 
}

\author{Arghadwip Paul}
\email{argha@umich.edu}
\affiliation{Department of Mechanical Engineering, University of Michigan, Ann Arbor, Michigan 48109, USA}

\author{Bikash Kanungo}
\affiliation{Department of Mechanical Engineering, University of Michigan, Ann Arbor, Michigan 48109, USA}

\author{Sambit Das}
\affiliation{Department of Mechanical Engineering, University of Michigan, Ann Arbor, Michigan 48109, USA}

\author{Vikram Gavini}
\email{vikramg@umich.edu}
\affiliation{Department of Mechanical Engineering, University of Michigan, Ann Arbor, Michigan 48109, USA}
\affiliation{Department of Materials Science and Engineering, University of Michigan, Ann Arbor, Michigan 48109, USA}

\date{\today}

\begin{abstract}
We present NNLap, a machine-learned Laplacian-level exchange-correlation (XC) functional that augments PBE with a neural-network correction depending on the electron density, its gradient, and its Laplacian.
The model is trained on exact XC potentials and energies, obtained through inverse density-functional theory (DFT) calculations on configuration-interaction densities.
Despite training on only a few systems -- five atoms and three molecules -- the model achieves remarkable accuracy on thermochemistry benchmarks, competing with the meta-GGA functionals SCAN and r2SCAN.
It also attains accurate total energies, comparable to SCAN and better than r2SCAN and B3LYP.
This shows that a Laplacian-level model, trained on exact XC potentials and energies, can reach the accuracy of meta-GGAs without their orbital dependence.
\end{abstract}

\maketitle

\section{\label{sec:Introduction} Introduction}

Density functional theory (DFT)~\cite{hohenberg1964inhomogeneous,kohn1965self} is the most widely used electronic-structure method in computational chemistry and materials science, owing to its favorable balance between accuracy and computational efficiency. Rather than solving the interacting many-electron Schr\"{o}dinger equation directly, DFT replaces it with an auxiliary noninteracting problem that reproduces the same electron density and is governed by an effective potential. Key to this simplification is the existence of a universal exchange–correlation (XC) energy functional ($\Exc$) that encodes the quantum many-electron interactions as a mean-field of the electron density ($\rho$).

Although DFT is formally exact, the exact XC functional is unknown and needs approximation. Finding better XC approximations has remained a central challenge over the past four decades, motivating the development of XC approximations. The existing approximations can, broadly, be classified into different rungs, often referred to as the Jacob’s ladder of DFT~\cite{perdew2001jacobs}. Moving up the ladder generally improves the accuracy of DFT, but also increases computational cost. These
approximations have been constructed by incorporating increasingly rich density-dependent information, together with exact constraints, known physical limits, and empirical or semi-empirical parameterization against reference data. At the lowest rung, the local density approximation (LDA) depends only on the electron density $\rho(\boldsymbol{r})$, while generalized gradient approximations (GGAs) additionally include the density gradient $\nabla \rho(\boldsymbol{r})$. Meta-GGAs further incorporate higher-level semilocal information, most commonly through the kinetic energy density $\tau(\boldsymbol{r})$, and hybrid functionals introduce a fraction of nonlocal exact exchange. This progression has produced widely used functionals such as PW92~\cite{perdew1992accurate}, PBE~\cite{perdew1996generalized}, SCAN~\cite{sun2015strongly}, r2SCAN~\cite{furness2020accurate}, B3LYP~\cite{becke1993density,lee1988development}, and $\omega$B97X-V~\cite{mardirossian2014wb97xv}. However, all existing XC approximations, remain far from the desired chemical accuracy of 1 kcal/mol in energies, which in turn severely limits the predictive powers of DFT. 

While the conventional XC approximations relied on human-designed forms, recently attempts have been to use machine-learning (ML) to design XC approximations, given the flexible functional forms that ML can offer~\cite{pederson2022machine,wu2023construct}. The earliest such attempt dates to Tozer and Handy's 1996 work~\cite{tozer1996exchange}, where it was limited by the ML tools and computing technology of its time. However, the rapid progress in both ML tools and computing technology in the last decade, has spurred active research in ML-based XC approximations. Such models can be constructed at a chosen rung of Jacob's ladder by
learning the dependence of the XC functional on the relevant density information
available at that level. This flexibility is attractive because it combines the adaptability of
empirical fitting with the physical grounding of non-empirical design. We can build
exact constraints and limiting behavior directly into the
ML architecture or impose them during training~\cite{dick2021highly,nagai2022constraints}. However, developing ML-based XC approximations has proven to be difficult. Recent attempts have tried to 
learned the XC potential, $\vxc[\rho](\br) = \frac{\Exc[\rho]}{\rho(\br)}$, instead of the XC energy functional, $\Exc[\rho]$~\cite{nagai2018neural}. 
But a potential learned in isolation need not be the functional
derivative of any energy functional, which undermines its use in
variational, self-consistent calculations. Later approaches restored
this consistency by learning the energy functional and training against
self-consistent densities and energies: by gradient-free
optimization~\cite{nagai2020completing}, by alternating parameter updates with
self-consistent solves~\cite{dick2020machine,chen2021deepks} or through differentiable Kohn--Sham
solvers~\cite{li2021kohnsham,kasim2021learning,dick2021highly}. In each case, a Kohn--Sham solution enters every training step, so the cost grows rapidly with the size of the model and of the training set.

In recent years, researchers have tried to improve accuracy by making the functional explicitly nonlocal, constructing it from nonlocal features of the density at an added computational cost.
Bystrom and Kozinsky learned exchange-only functionals for molecules and solids from exact-exchange energies, using nonlocal density features that build in the uniform coordinate scaling of exchange~\cite{bystrom2022cider,bystrom2024nonlocal}.
Riemelmoser \textit{et al.} built a nonlocal extension of the GGA from two- and three-body descriptors of the density, using reference data from the random-phase approximation~\cite{riemelmoser2023machine}. However, these models are fitted using training sets that are similar to their test sets, and hence, their out-of-distribution accuracy remains uncertain. Some more sophisticated models have adopted the strategy of scaling up the training data. DM21~\cite{kirkpatrick2021pushing} uses a flexible local-hybrid form with $\sim$4000,000
parameters trained on over thousand molecules. It
attains impressive accuracy on the chemistry it is trained on. But
questions about transferability beyond that data, and about robustness
in self-consistent calculations, remain open~\cite{gerasimov2022comment,kirkpatrick2022response}. Skala~\cite{luise2025accurate} is a more recent nonlocal model with $\sim$385,000 parameters trained on a massive dataset of $\sim$150,000 molecules. However, it captures nonlocality through density features centered on each nuclei. As a result, it deviates from being a density functional, as it inadvertently involves geometry information.     


Kanungo \textit{et al.}~\cite{kanungo2025learning} took a different route: rather than using more training data, they made each training system
more informative by using the information of the exact XC potential, in addition to exact density and XC energies. The exact XC potentials and energies were obtained using their recently developed \textit{inverse} DFT method \textit{inverse} DFT~\cite{kanungo2019exact, Kanungo2023} on full configuration-interaction quality densities. 
The XC potential is the
key ingredient: an energy contributes one number per system, whereas
the potential constrains the functional derivative of the model at
every point in space. Training against it requires no self-consistent
calculations, so the earlier cost barrier does not arise.
Trained on just a handful of atoms and small molecules, their neural-network (NN) based GGA functional, hereafter referred as NNGGA-UEG,
improved total energies and densities well outside the training
set on thermochemistry benchmarks.

In this work, we build on the work of Kanungo \textit{et al.} and extend their semilocal form beyond the
GGA level by adding the Laplacian of the electron density,
$\nabla^2 \rho(\boldsymbol{r})$, as a descriptor in the NN model. $\nabla^2 \rho$ provides local curvature information beyond $\rho$ and $\nabla\rho$ used in GGA, while remaining an explicit density-dependent quantity within the conventional Kohn-Sham DFT framework. We term this model as NNLap. NNLap offers a cost-effective route to enrich the functional form without introducing the orbital dependence associated with meta-GGAs (which use kinetic-energy density ($\tau$) based) or the hybrid functionals (which use a fraction of the exact exchange).

The NNLap, trained on just five atoms and three molecules, attains a remarkable accuracy of 2.3 kcal/mol on total energies per atom, surpassing the SCAN and the r2SCAN meta-GGAs, and the B3LYP hybrid functionals, which have nonlocal information through their orbital dependence.  
On the W4-11 atomization energies, the mean absolute error drops from 14.962~kcal/mol for PBE and 9.357~kcal/mol for the gradient-only NNGGA-UEG to 6.477~kcal/mol for NNLap.
On 19 dispersion-free subsets of GMTKN55~\cite{goerigk2017gmtkn55}, NNLap attains similar accuracy as the SCAN and r2SCAN meta-GGAs.

\section{Theory and Methods}
\label{sec:Theory}

\subsection{Forward and Inverse Kohn-Sham DFT}

\begin{figure*}[t]
\centering
\resizebox{\textwidth}{!}{%
\begin{tikzpicture}[
    >=Latex,
    font=\large,
    box/.style={
        draw,
        rounded corners=4pt,
        line width=0.9pt,
        minimum height=4.3cm,
        align=center
    },
    smallbox/.style={
        draw=blue!80!black,
        rounded corners=4pt,
        line width=0.9pt,
        minimum width=1.6cm,
        minimum height=1.25cm,
        align=center
    },
    flow/.style={
        -{Latex[length=3mm,width=2mm]},
        line width=1.0pt
    }
]

\node[
    box,
    minimum width=4.0cm
] (mol) {};

\node[
    anchor=north,
    font=\bfseries\large
] at ([yshift=-4mm]mol.north)
{Training system};

\node at ([yshift=-2mm]mol.center)
{\includegraphics[width=2.8cm]{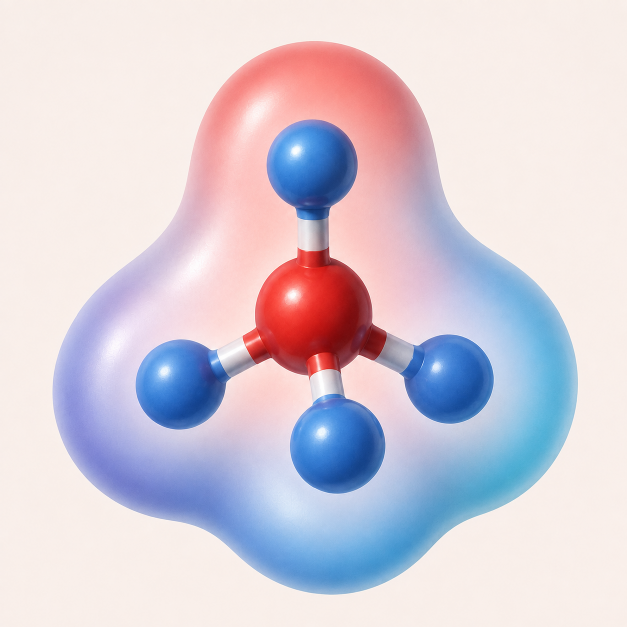}};

\node[
    box,
    draw=red,
    minimum width=3.1cm,
    right=0.55cm of mol
] (ci) {};

\node[
    anchor=north,
    text=red,
    font=\bfseries\Large
] at ([yshift=-5mm]ci.north)
{CI};

\node[font=\normalsize] at ([yshift=3mm]ci.center)
{$\hat{H}\Psi = E\Psi$};

\node[font=\normalsize] at ([yshift=-8mm]ci.center)
{$
|\Psi_{\mathrm{CI}}\rangle
=
\displaystyle\sum_I c_I |\Phi_I\rangle
$};

\draw[flow] (mol.east) -- (ci.west);

\node[
    box,
    draw=blue!80!black,
    minimum width=3.2cm,
    right=1.8cm of ci
] (idft) {};

\node[
    anchor=north,
    text=blue!80!black,
    font=\bfseries\Large
] at ([yshift=-5mm]idft.north)
{Inverse DFT};

\node[
    align=center,
    font=\small
] (rho) at ($(ci.east)!0.5!(idft.west)+(0,0.6)$)
{Reference\\[-1mm]$\rho(\mathbf r),\ E$};

\draw[flow] (ci.east) -- (idft.west);

\node at ([yshift=-4mm]idft.center)
{\includegraphics[width=3.0cm]{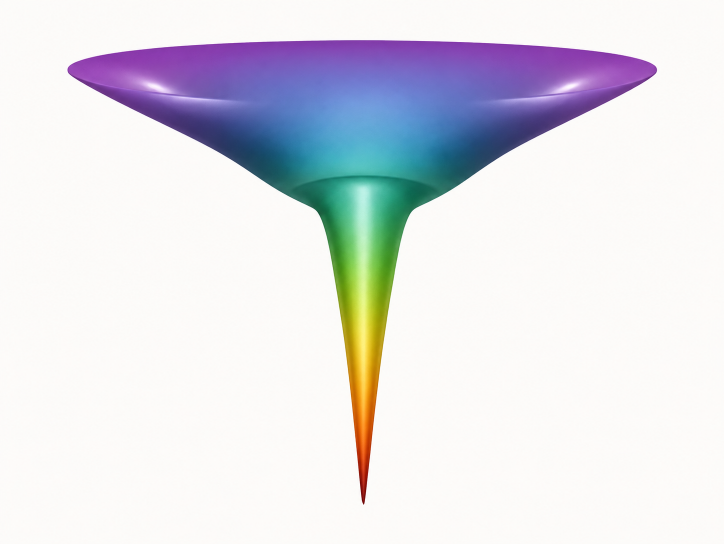}};

\node at ([yshift=-18mm]idft.center)
{$v_{\rm xc}(\mathbf r)$};

\node[
    smallbox,
    right=0.7cm of idft,
    yshift=1.05cm
] (exc)
{$E_{\rm xc}$};

\node[
    smallbox,
    right=0.7cm of idft,
    yshift=-1.05cm
] (vxc)
{$v_{\rm xc}(\mathbf r)$};

\node[
    above=4mm of exc,
    text=blue!80!black,
    font=\bfseries\large,
    align=center
]
{Training\\targets};

\draw[flow] ([yshift=1.05cm]idft.east) -- (exc.west);
\draw[flow] ([yshift=-1.05cm]idft.east) -- (vxc.west);

\node[
    box,
    draw=green!30!black,
    minimum width=5.0cm,
    right=1.05cm of exc,
    yshift=-1.05cm
] (dnnbox) {};

\node[
    anchor=north,
    text=green!30!black,
    font=\bfseries\Large
] at ([yshift=-4mm]dnnbox.north)
{DNN};

\coordinate (in1) at ([xshift=1.35cm,yshift= 0.9cm]dnnbox.west);
\coordinate (in2) at ([xshift=1.35cm,yshift= 0.3cm]dnnbox.west);
\coordinate (in3) at ([xshift=1.35cm,yshift=-0.3cm]dnnbox.west);
\coordinate (in4) at ([xshift=1.35cm,yshift=-0.9cm]dnnbox.west);

\node[left=2mm of in1, font=\small] {$\rho_{\uparrow}(\mathbf r)$};
\node[left=2mm of in2, font=\small] {$\rho_{\downarrow}(\mathbf r)$};
\node[left=2mm of in3, font=\small] {$\nabla\rho(\mathbf r)$};
\node[left=2mm of in4, font=\small] {$\nabla^2\rho(\mathbf r)$};

\tikzset{
    neuron/.style={
        circle,
        draw=black,
        fill=green!65,
        minimum size=4mm,
        inner sep=0pt
    }
}

\node[neuron] (i1) at ([xshift=1.75cm,yshift= 0.9cm]dnnbox.west) {};
\node[neuron] (i2) at ([xshift=1.75cm,yshift= 0.3cm]dnnbox.west) {};
\node[neuron] (i3) at ([xshift=1.75cm,yshift=-0.3cm]dnnbox.west) {};
\node[neuron] (i4) at ([xshift=1.75cm,yshift=-0.9cm]dnnbox.west) {};

\foreach \y/\n in {0.95/h11,0.48/h12,0/h13,-0.48/h14,-0.95/h15}
    \node[neuron] (\n) at ([xshift=2.7cm,yshift=\y cm]dnnbox.west) {};

\foreach \y/\n in {0.9/h21,0.3/h22,-0.3/h23,-0.9/h24}
    \node[neuron] (\n) at ([xshift=3.7cm,yshift=\y cm]dnnbox.west) {};

\node[neuron] (out) at ([xshift=4.55cm]dnnbox.west) {};

\foreach \i in {i1,i2,i3,i4}{
    \foreach \h in {h11,h12,h13,h14,h15}{
        \draw[thin] (\i) -- (\h);
    }
}

\foreach \h in {h11,h12,h13,h14,h15}{
    \foreach \g in {h21,h22,h23,h24}{
        \draw[thin] (\h) -- (\g);
    }
}

\foreach \g in {h21,h22,h23,h24}{
    \draw[thin] (\g) -- (out);
}

\draw[flow] (exc.east) -- ([yshift=0.9cm]dnnbox.west);
\draw[flow] (vxc.east) -- ([yshift=-0.9cm]dnnbox.west);

\node[
    box,
    draw=violet,
    minimum width=2.8cm,
    right=0.55cm of dnnbox
] (output) {};

\node[
    anchor=north,
    text=violet,
    font=\bfseries\Large
] at ([yshift=-5mm]output.north)
{Output};

\node[
    text=violet,
    align=center
] at ([yshift=4mm]output.center)
{Learned\\XC functional};

\node at ([yshift=-8mm]output.center)
{$E_{\rm xc}[\rho]$};

\draw
([xshift=4mm,yshift=-3mm]output.west |- output.center)
--
([xshift=-4mm,yshift=-3mm]output.east |- output.center);

\node at ([yshift=-16mm]output.center)
{$
v_{\rm xc}(\mathbf r)
=
\dfrac{\delta E_{\rm xc}}
      {\delta \rho(\mathbf r)}
$};

\draw[flow] (dnnbox.east) -- (output.west);

\node[
    draw=green!30!black,
    rounded corners=4pt,
    line width=0.9pt,
    minimum width=6.0cm,
    minimum height=1.45cm,
    below=0.35cm of idft,
    xshift=1.0cm,
    align=center
] (desc)
{
{\bfseries\color{green!30!black}Density descriptors}
\\[1mm]
$\rho_{\uparrow}(\mathbf r),\;
 \rho_{\downarrow}(\mathbf r),\;
 \nabla\rho(\mathbf r),\;
 \nabla^2\rho(\mathbf r)$
};

\draw[line width=1pt]
(ci.south)
|- (desc.west);

\draw[flow]
(desc.east)
-| ([xshift=-0.55cm]dnnbox.south);

\end{tikzpicture}%
}
\caption{
Workflow for constructing the NN-based XC functional.
CI calculations on the training systems provide the exact reference densities and total energies, which inverse DFT uses to calculate the corresponding exact XC potentials and XC energies.
The NN uses the densities, XC potentials, and XC energies to model the XC functional.
}
\label{fig:training_workflow}
\end{figure*}
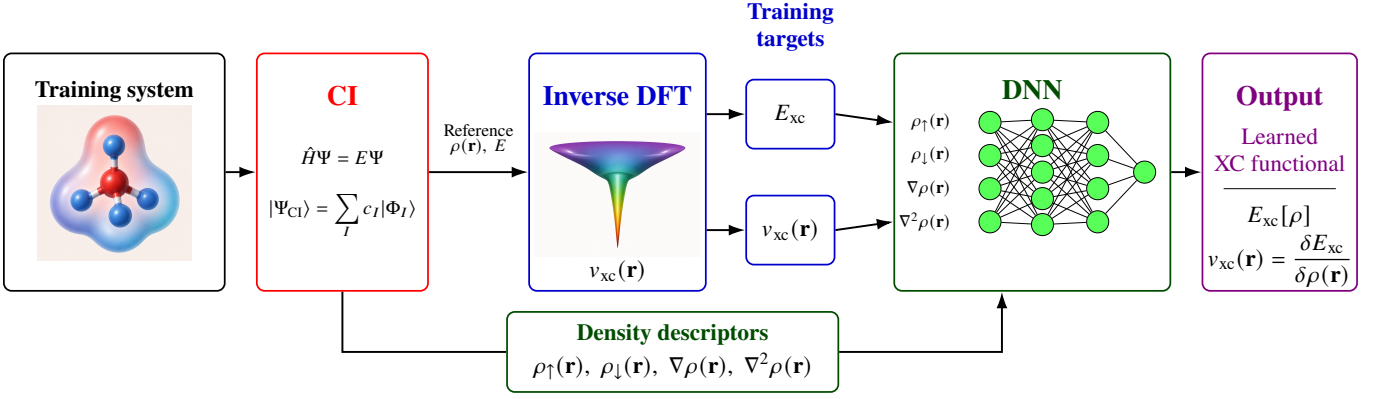

We train our model on exact XC potentials and energies, both obtained by inverse Kohn-Sham (KS) DFT. Fig.~\ref{fig:training_workflow} provides a schematic of our work.
We first summarize the forward KS problem and then pose the inverse problem.

In spin-unrestricted Kohn-Sham density functional theory, the ground-state spin densities
\(\rho^\sigma(\mathbf r)\), with \(\sigma \in \{\alpha,\beta\}\), are obtained from an auxiliary noninteracting system. The KS orbitals \(\phi_i^\sigma(\mathbf r)\) and their eigenvalues \(\epsilon_i^\sigma\) satisfy
\begin{equation}
\left[
-\frac{1}{2}\nabla^2
+
v_s^\sigma(\mathbf r)
\right]
\phi_i^\sigma(\mathbf r)
=
\epsilon_i^\sigma \phi_i^\sigma(\mathbf r).
\label{eq:ks-equation}
\end{equation}
The spin densities are obtained from the orbitals as
\begin{equation}
\rho^\sigma(\mathbf r)
=
\sum_i
f_i^\sigma
\left|\phi_i^\sigma(\mathbf r)\right|^2,
\qquad
\rho(\mathbf r)=\rho^\alpha(\mathbf r)+\rho^\beta(\mathbf r),
\label{eq:spin-density}
\end{equation}
where \(f_i^\sigma\) is the occupancy of the orbital \(\phi_i^\sigma\), often determined through Fermi-Dirac statistics.
The effective KS potential is decomposed as
\begin{equation}
v_s^\sigma(\mathbf r)
=
v_{\rm ext}(\mathbf r)
+
v_{\rm H}[\rho](\mathbf r)
+
v_{\rm xc}^\sigma[\rho^\alpha,\rho^\beta](\mathbf r),
\label{eq:ks-potential}
\end{equation}
where \(v_{\rm ext}\), \(v_{\rm H}\), and \(v_{\rm xc}^\sigma\) are the external,
Hartree, and exchange-correlation (XC) potentials, respectively.
The XC potential is obtained from the XC energy functional \(E_{\rm xc}[\rho^\alpha,\rho^\beta]\) through the functional derivative
\begin{equation}
v_{\rm xc}^\sigma(\mathbf r)
=
\frac{\delta E_{\rm xc}[\rho^\alpha,\rho^\beta]}
{\delta \rho^\sigma(\mathbf r)} .
\label{eq:vxc-functional-derivative}
\end{equation}

In inverse KS theory, the direction of this mapping is reversed. Given accurate
reference spin densities \(\rho_{\rm ref}^\sigma(\mathbf r)\), obtained here from configuration interaction (CI) calculations~\cite{chien2018excited,dang2022slater,dang2023advances}, one determines 
spin-dependent KS potentials \(v_{s,{\rm ref}}^\sigma(\mathbf r)\) whose
noninteracting ground state reproduces the reference densities.
This inverse problem is posed as an optimization over trial KS
potentials,
\begin{equation}
\begin{split}
\{v_{s,{\rm ref}}^\alpha,v_{s,{\rm ref}}^\beta\}
=
\arg\min_{\{\tilde v_s^\alpha,\tilde v_s^\beta\}}
&\sum_\sigma
\int
w(\mathbf r)
\\
&\times
\left|
\rho_{\tilde v_s}^\sigma(\mathbf r)
-
\rho_{\rm ref}^\sigma(\mathbf r)
\right|^2
\,d\mathbf r ,
\end{split}
\label{eq:inverse-ks-optimization}
\end{equation}
subject to the KS equations in Eq.~\eqref{eq:ks-equation} and the orthonormality of the orbitals.
Here, \(\rho_{\tilde v_s}^\sigma\) is the spin density obtained by solving Eq.~\eqref{eq:ks-equation} with the trial potentials \(\{\tilde v_s^\alpha,\tilde v_s^\beta\}\), and \(w(\mathbf r)\) is a positive weight chosen to improve convergence in low-density regions.
We solve this optimization using the finite-element based inverse DFT method of Kanungo \textit{et al.}~\cite{kanungo2019exact, Kanungo2023, subramanian2026invdft}.

Once the reference KS potentials are obtained, the corresponding XC potentials
are extracted from the KS decomposition,
\begin{equation}
v_{\rm xc,ref}^\sigma(\mathbf r)
=
v_{s,{\rm ref}}^\sigma(\mathbf r)
-
v_{\rm ext}(\mathbf r)
-
v_{\rm H}[\rho_{\rm ref}](\mathbf r),
\label{eq:inverse-vxc}
\end{equation}
where \(\rho_{\rm ref}=\rho_{\rm ref}^\alpha+\rho_{\rm ref}^\beta\) is the total reference density.
The corresponding XC energy follows from the KS decomposition of the reference total energy \(E_{\rm ref}\), taken from the same CI calculation,
\begin{equation}
E_{\rm xc}
=
E_{\rm ref}
-
\sum_{\sigma}
T_s[\rho_{\rm ref}^\sigma]
-
E_{\rm H}[\rho_{\rm ref}]
-
\int
\rho_{\rm ref}(\mathbf r)\,
v_{\rm ext}(\mathbf r)
\,d\mathbf r .
\label{eq:inverse-exc}
\end{equation}
Here, \(T_s[\rho_{\rm ref}^\sigma]=\frac{1}{2}\sum_i f_i^\sigma \int |\nabla \phi_i^\sigma(\mathbf r)|^2 \,d\mathbf r\) is the noninteracting kinetic energy, evaluated using the KS orbitals obtained from the inverse solution.
\(E_{\rm H}[\rho_{\rm ref}]\) is the Hartree energy of the reference density.
Thus, the inverse KS procedure provides with $\{\Exc^{\mathrm{ref}}, v_{\mathrm{xc,ref}}^{\sigma}\}$ as high fidelity training data to machine-learn the XC energy functional ($\Exc[\rho^{\alpha}, \rho^{\beta}]$).

\subsection{NN-based XC functional}
\label{sec:nn-functional}

We construct the NNLap XC functional as an NN-based extension to the PBE functional. That is, instead of using an NN to represent the full XC energy, we use PBE as a base semilocal approximation and use the NN to describe an additive correction. This provides the model with a physically motivated baseline while allowing the network to learn features of the reference XC energies and potentials that are not adequately represented by PBE. For simplicity, hereafter in equations, we use the superscript NN as a shorthand for NNLap.

The XC energy is written as
\begin{equation}
E_{\rm xc}^{\rm NN}
=
\int
e_{\rm xc}^{\rm NN}[\rho^\alpha, \rho^\beta, |\nabla \rho|, \nabla^2\rho](\mathbf{r})
\,d\mathbf{r},
\label{eq:pbe-nn-energy}
\end{equation}
with the local XC energy density
\begin{equation}
\begin{split}
e_{\rm xc}^{\rm NN}[\rho^\alpha, \rho^\beta, |\nabla \rho|, \nabla^2\rho]
&=
e_{\rm xc}^{\rm PBE}[\rho^\alpha, \rho^\beta, |\nabla \rho|]
~+ \\
&e_x^{\rm UEG}(\rho)\,
\phi(\xi)\,
\tanh(s^2)\,
G_{\boldsymbol{\theta}}(\mathbf{x}).
\end{split}
\label{eq:pbe-nn-model}
\end{equation}
Here, $e_{\rm xc}^{\rm PBE}$ is the PBE XC energy density and
$G_{\boldsymbol{\theta}}$ is a feed-forward neural network
with inputs $\mathbf{x}$ (see Eq.~\eqref{eq:nn-x-map}) and trainable parameters $\boldsymbol{\theta}$. In the above, $\rho = \rho^\alpha + \rho^\beta$. 
The prefactors $e_x^{\rm UEG}(\rho)$ and $\phi(\xi)$ set density and spin dependence of the dominant exchange part of the XC energy. The factor $\tanh(s^2)$ is used to recover the PBE functional in the slowly varying density limit (i.e.,$s\rightarrow0$). 
We define each of the factors and their inputs below.

$\xi = \frac{\rho^\alpha-\rho^\beta}{\rho}$ is the relative spin polarization is defined as
The spatial variation of $\rho$ is described by the reduced
density gradient and reduced Laplacian,
\begin{equation}
s
=
\frac{|\nabla\rho|}
{2(3\pi^2)^{1/3}\rho^{4/3}},
\qquad
q
=
\frac{\nabla^2\rho}
{4(3\pi^2)^{2/3}\rho^{5/3}}.
\label{eq:nn-s-q}
\end{equation}
While $s$ measures the magnitude of the local density variation,
$q$ characterizes the local curvature of the
density. $e_x^{\rm UEG}(\rho)
= -\frac{3}{4}
\left(\frac{3}{\pi}\right)^{1/3}
\rho^{4/3}$ is the exchange energy density of the uniform electron gas (UEG). $\phi(\xi) =  \frac{1}{2}
\left(
(1+\xi)^{4/3}
+
(1-\xi)^{4/3}
\right)$ is the spin-scaling factor. The input $\textbf{x}$ to $G_{\theta}$ are transformations of $\rho$, $\xi$, $s$, and $q$, to ensure that, quantities that otherwise span different order of magnitudes, are mapped to similar numerical ranges. Specifically, $\mathbf{x} =
\left(x_\rho, x_\xi, x_s,
x_q \right),$ where 
\begin{equation}
    \begin{split}
    x_\rho &= \ln\left(\rho\right),\,\,\,
    x_\xi = \xi^2,\\ 
    x_s &= \ln \left( 1+s \right),\,\,\,
    x_q = \operatorname{asinh}\left(\frac{q}{5}\right).
\label{eq:nn-x-map}
\end{split}
\end{equation}
The logarithmic transformations are used to compress the large dynamic ranges of $\rho$ and $s$. The transformation
$\operatorname{asinh}(q/5)$ compresses large positive and negative values
of the reduced Laplacian while preserving its sign and remaining
approximately linear near $q=0$.
The scale factor of 5 keeps energetically important values of $q$ within the near-linear range of asinh function. Finally, the use of $\xi^2$ ensures that
the XC energy is invariant under interchange of the two spin channels.

Because the neural network is introduced at the energy-functional level,
the corresponding spin-dependent XC potentials are obtained by functional
differentiation of the complete NNLap energy,
\begin{equation}
v_{\rm xc}^{\sigma,{\rm NN}}(\mathbf{r})
=
\frac{
\delta E_{\rm xc}^{\rm NN}
}{
\delta\rho^\sigma(\mathbf{r})
}.
\label{eq:nn-xc-potential}
\end{equation}
The XC energies and potentials used during training are therefore generated
from the same underlying energy functional.

\subsection{NN training}
\label{sec:nn-training}

Our network $G_{\boldsymbol{\theta}}$ consists of three fully connected hidden layers with 80 neurons per layer. The exponential linear unit (ELU) is used as the nonlinear activation function in each hidden layer, while the output layer is linear. The network parameters $\boldsymbol{\theta}$ comprise the weights and biases of all layers.

The training set comprises five atoms (Li, C, N, O, and Ne) and three molecules (H$_2$, LiH, and triplet CH$_2$).
Following Kanungo \textit{et al.}~\cite{kanungo2025learning}, we also include the energy and potential of the UEG as additional training data.
The network is trained using the reference XC energies and spin-dependent XC potentials of these systems, evaluated at the reference densities. We use a composite loss ($\mathcal{L}$), comprising an energy loss ($\mathcal{L}_E$) and a potential loss term ($\mathcal{L}_v$), given as 
\begin{equation}
\mathcal{L}(\boldsymbol{\theta})
=
c_E\mathcal{L}_{E}
+
\mathcal{L}_{v},
\label{eq:nn-total-loss}
\end{equation}
where $c_E$ controls the relative weight of the energy contribution.
We set $c_E=10^{6}$, which brings the energy and potential losses to a comparable magnitude. The energy loss ($\mathcal{L}_E$) is defined as
\begin{equation}
\mathcal{L}_{E}
=
\frac{1}{N_{\rm sys}}
\sum_{I=1}^{N_{\rm sys}}
\left(
E_{{\rm xc},I}^{\rm NN}
-
E_{{\rm xc},I}^{\rm ref}
\right)^2,
\label{eq:nn-energy-loss}
\end{equation}
where $I$ indexes the training systems and $E_{{\rm xc},I}^{\rm NN}$ is the energy of Eq.~\eqref{eq:pbe-nn-energy} evaluated at the reference densities of system $I$. The potential loss ($\mathcal{L}_v$) is defined as
\begin{equation}
\begin{split}
\mathcal{L}_{v}
=
\frac{1}{N_{\rm sys}}
\sum_{I=1}^{N_{\rm sys}}
\sum_{\sigma}
\biggl[
\int
\rho_I^\sigma(\mathbf{r})
\Bigl(
&v_{{\rm xc},I}^{\sigma,{\rm NN}}(\mathbf{r}) -
v_{{\rm xc},I}^{\sigma,{\rm ref}}(\mathbf{r})
\Bigr)
\,d\mathbf{r}
\biggr]^2.
\end{split}
\label{eq:nn-potential-loss}
\end{equation}
Here, $v_{{\rm xc},I}^{\sigma,{\rm NN}}$ is the model XC potential of Eq.~\eqref{eq:nn-xc-potential} evaluated at the reference densities of system $I$. In the above, we use the  density weighting to weight the energetically relevant regions more as well as provide both the loss terms the dimension of the square of the energy.
Since $e_{\rm xc}^{\rm NN}$ depends on $\rho^\alpha$, $\rho^\beta$, $\nabla\rho$, and $\nabla^2\rho$, carrying out the functional derivative in Eq.~\eqref{eq:nn-xc-potential} gives
\begin{equation}
\begin{split}
v_{\rm xc}^{\sigma,{\rm NN}}
=
\frac{\partial e_{\rm xc}^{\rm NN}}{\partial \rho^\sigma}
&-
\nabla\cdot
\frac{\partial e_{\rm xc}^{\rm NN}}{\partial(\nabla\rho)}
+
\nabla^2\,
\frac{\partial e_{\rm xc}^{\rm NN}}{\partial(\nabla^2\rho)}.
\end{split}
\label{eq:nn-vxc-explicit}
\end{equation}
We compute the partial derivatives of $e_{\rm xc}^{\rm NN}$ with respect to its arguments using automatic differentiation.
Evaluating the outer spatial derivatives the same way would require nested, higher-order automatic differentiation, which is computationally costly and difficult to implement correctly~\cite{baydin2018automatic,griewank2008evaluating}.
We therefore evaluate the divergence and the Laplacian terms in  Eq.~\eqref{eq:nn-vxc-explicit} by finite difference.

The NN is implemented via the Pytorch software package~\cite{Paszke2019}. The weights and biases in the NN are optimized with Adam optimizer using a learning rate of $10^{-3}$.
We stop training once the loss plateaus, that is, when its standard deviation over 100 consecutive epochs falls below $10^{-3}$. After training the NNLap model, we integrate it with the PySCF quantum chemistry package~\cite{Sun2018, Sun2020} to perform the Kohn-Sham self-consistent field (SCF) calculations on the test systems. We refer to Appendix~\ref{app:scf} for the details on the SCF calculations. 


\section{\label{sec:Results} Results}

\subsection{Enhancement-factor analysis}

\begin{figure*}[t]
    \centering
    \includegraphics[width=\textwidth]{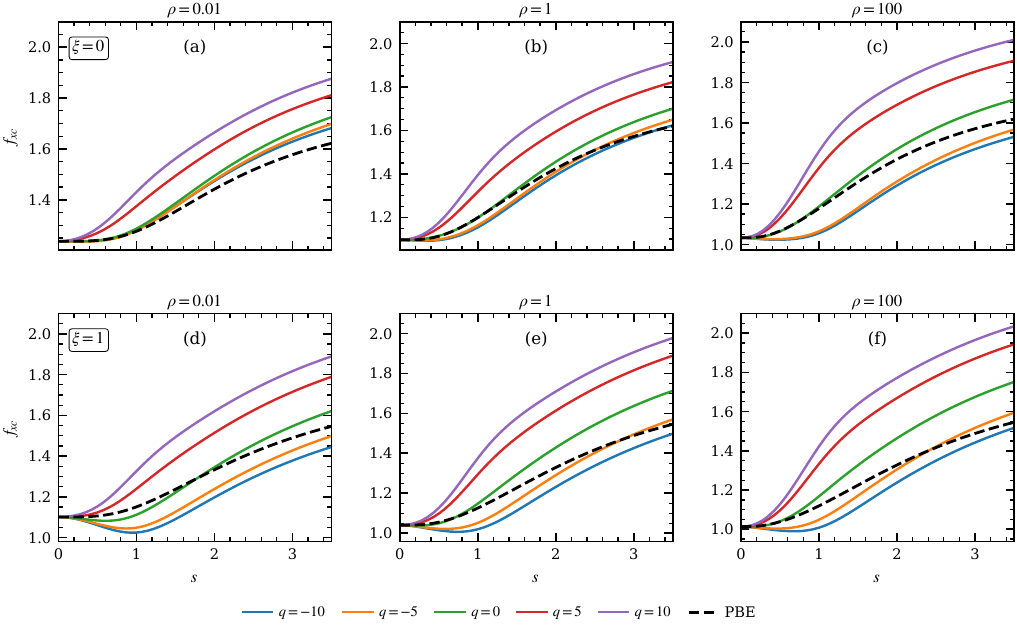}
    \caption{Enhancement factor $F_{\mathrm{xc}}$ of NNLap as a function of the reduced density gradient $s$, for reduced Laplacian values $q=-10$, $-5$, $0$, $5$, and $10$ (solid curves).
    Panels (a)--(c) correspond to $\xi=0$ and panels (d)--(f) to $\xi=1$, for densities $\rho=0.01$, $1$, and $100$, respectively.
    The dashed curve in each panel shows the PBE enhancement factor.}
    \label{fig:fxc_vs_s}
\end{figure*}

To examine the local behavior learned by NNLap, we consider the
XC enhancement factor
\begin{equation}
F_{\mathrm{xc}}(\rho,\xi,s,q)
=
\frac{
e_{\mathrm{xc}}^{\mathrm{NN}}(\rho,\xi,s,q)
}{
e_x^{\mathrm{UEG}}(\rho)
},
\end{equation}
shown in Fig.~\ref{fig:fxc_vs_s} as a function of the reduced
density gradient \(s\) for representative values of the density
\(\rho\), spin polarization \(\xi\), and reduced Laplacian \(q\).
Because the NN correction is multiplied by \(\tanh(s^2)\), all
NNLap curves recover the underlying PBE enhancement factor at
\(s=0\), independently of \(q\). In particular, the \(s=0\), \(q=0\)
limit corresponds to the uniform electron gas. The smooth approach to
PBE near this limit avoids the sharp small-\(s\) variation observed by
Kanungo \textit{et al.}~\cite{kanungo2025learning} for an unconstrained NNGGA model and retains the
known uniform-density behavior of the PBE baseline.

Away from the uniform-density limit, the curves develop a pronounced
dependence on \(q\). For fixed \(\rho\), \(\xi\), and \(s\), increasing
\(q\) systematically increases \(F_{\mathrm{xc}}\), whereas negative
values of \(q\) reduce it. The separation between the different
\(q\)-dependent curves is small near \(s=0\) and becomes progressively
larger with increasing \(s\). Thus, the Laplacian dependence is
suppressed in the slowly varying limit but becomes significant in more
strongly inhomogeneous density regions. NNLap can therefore distinguish local
environments with the same density and gradient magnitude but
different density curvature, showing the additional descriptive
flexibility provided by the reduced Laplacian.

Across the range displayed in Fig.~\ref{fig:fxc_vs_s},
\(F_{\mathrm{xc}}\) remains smooth and below the
Lieb--Oxford-based value of \(2.215\)~\cite{lieb1981improved,chan1999optimized}, with no visible pathological
growth or oscillatory behavior. These plots therefore show that the
network introduces a substantial and structured Laplacian dependence
while retaining the PBE uniform-electron-gas limit.

This structured Laplacian dependence has a natural real-space interpretation.
Regions of charge concentration, such as covalent bonds and atomic shells, have $\nabla^2\rho < 0$ and hence $q < 0$, whereas density tails and depletion regions have $\nabla^2\rho > 0$ ($q > 0$)~\cite{bader1984characterization}.
Figure~\ref{fig:fxc_vs_s} then implies that, relative to PBE, NNLap lowers the enhancement factor in charge-concentration regions and raises it in tail and depletion regions.
Atoms and molecules weight these regions differently, so this distinction can shift the balance between atomic and molecular energies that atomization energies probe.
We note one caveat: exactly at a bond critical point the density gradient vanishes, so the $\tanh(s^2)$ factor suppresses the correction there regardless of $q$; the discrimination operates in the surrounding bonding region, where $s$ is small but nonzero.

\subsection{Thermochemistry tests}

\begin{figure*}[t]
    \centering
    \includegraphics[width=\textwidth]{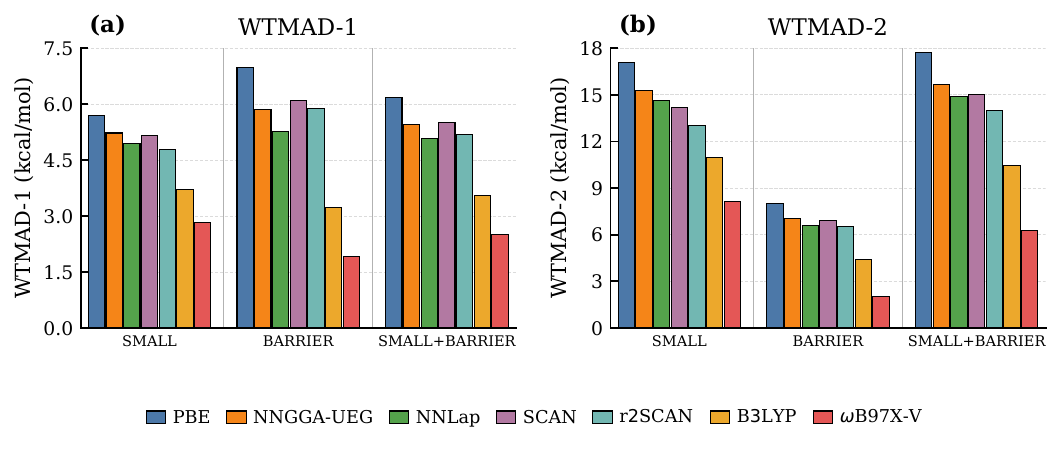}
    \caption{
    WTMAD-1 and WTMAD-2 errors for the SMALL, BARRIER, and combined SMALL+BARRIER benchmark groups.
    Panel (a) shows WTMAD-1 and panel (b) shows WTMAD-2. 
    The same functional ordering and color scheme are used in both panels.
    }
    \label{fig:wtmad_two_panel}
\end{figure*}

We assess the thermochemical performance of our functional using the W4-11~\cite{karton2011w4} and BH76~\cite{zhao2005htbh,zhao2005nhtbh} benchmark sets. W4-11 comprises 140 total atomization energies of small first- and second-row molecules and radicals, whereas BH76 contains 76 forward and reverse reaction barrier heights. Together, these datasets probe complementary bonding regimes: W4-11 tests the description of molecular bonding and dissociation across diverse electronic structures, while BH76 tests the relative energies of transition states with respect to the corresponding reactants and products.
Among the functionals compared, NNGGA-UEG is the gradient-level model of Kanungo \textit{et al.}~\cite{kanungo2025learning}, trained within the same framework.
Comparing against it isolates the effect of adding the reduced Laplacian.
The resulting mean errors (ME) and mean absolute errors (MAE) are summarized in Table~\ref{tab:w411_bh76}.

\begin{table}[h]
\centering
\caption{Mean errors (ME) and mean absolute errors (MAE), in
kcal/mol, for the W4-11 and BH76 benchmark sets.}
\label{tab:w411_bh76}

\begin{tabular*}{\columnwidth}{@{\extracolsep{\fill}}lrrrr@{}}
\toprule
& \multicolumn{2}{c}{W4-11}
& \multicolumn{2}{c}{BH76} \\
\cmidrule(lr){2-3}
\cmidrule(lr){4-5}
Functional & ME & MAE & ME & MAE \\
\midrule
PBE             & 13.355 & 14.962 & -9.108 & 9.151 \\
NNGGA-UEG       &  4.665 &  9.357 & -8.397 & 8.421 \\
NNLap       & -0.762 &  6.477 & -7.939 & 8.001 \\
SCAN            & -0.174 &  4.005 & -7.358 & 7.657 \\
r2SCAN       &  0.890 &  3.829 & -7.185 & 7.232 \\
B3LYP           & -2.951 &  4.216 & -4.019 & 4.936 \\
$\omega$B97X-V  & -0.795 &  2.781 & -1.215 & 1.832 \\
\bottomrule
\end{tabular*}
\end{table}

For W4-11, PBE strongly overestimates the atomization energies, as indicated by its large positive ME of 13.355 kcal/mol. This systematic bias is substantially reduced by NNGGA-UEG and is nearly eliminated by NNLap, for which the ME decreases to -0.762 kcal/mol. The inclusion of the reduced Laplacian also lowers the MAE from 14.962 kcal/mol for PBE and 9.357 kcal/mol for the gradient-only NNGGA-UEG model to 6.477 kcal/mol. Thus, the improvement is not merely due to cancellation between positive and negative errors, but is accompanied by a meaningful reduction in the typical magnitude of the individual errors. SCAN and r$^{2}$SCAN yield lower MAEs of 4.005 and 3.829 kcal/mol, respectively, showing the additional accuracy obtained from the kinetic-energy-density dependence of these meta-GGAs. B3LYP gives a comparable, though slightly larger, MAE of 4.216 kcal/mol, whereas the range-separated hybrid $\omega$B97X-V provides the lowest error, with an MAE of 2.781 kcal/mol. Overall, including the reduced Laplacian in our functional leads to a substantial improvement over PBE for atomization energies.

The behavior is qualitatively different for BH76. Reaction barriers involve transition-state geometries with stretched, weak bonds, for which nonlocal exchange effects are particularly important. Accordingly, all semilocal functionals considered here systematically underestimate the barrier heights, as reflected by their large negative MEs. NNGGA-UEG and NNLap reduce the MAE of PBE from 9.151 to 8.421 and 8.001 kcal/mol, respectively, but the improvement is considerably smaller than for W4-11. SCAN and r$^{2}$SCAN similarly retain substantial MAEs of 7.657 and 7.232 kcal/mol. In contrast, the hybrid functionals perform markedly better, with MAEs of 4.936 kcal/mol for B3LYP and 1.832 kcal/mol for $\omega$B97X-V. These results indicate that additional semilocal density information alone provides only limited improvement for reaction barriers, whereas nonlocal ingredients, particularly exact exchange, are needed for a substantial reduction in the error.

\begin{figure*}[t]
    \centering
    \includegraphics[width=\textwidth]{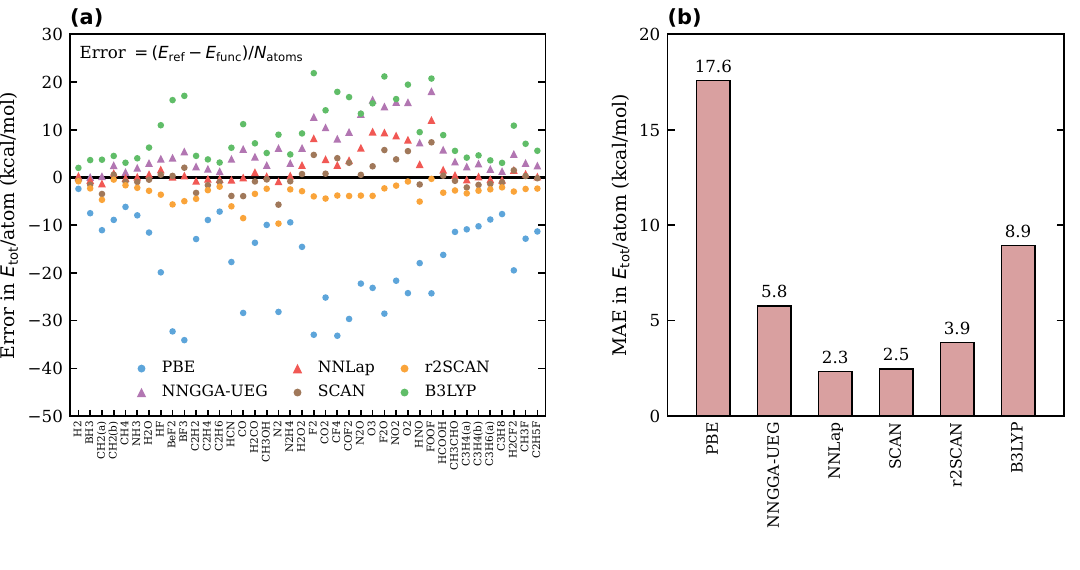}
    \caption{
    Errors in the total energy per atom, $(E_{\mathrm{ref}} - E_{\mathrm{func}})/N_{\mathrm{atoms}}$, in kcal/mol.
    (a) Errors for some molecules of the W4-11 dataset.
    Molecules with a parenthesis indicate different isomers or spin states.
    (b) MAE for the 99 molecules in W4-11 containing up to second-row elements.
    }
    \label{fig:total_energy}
\end{figure*}

\begin{table*}[t]
\centering
\caption{Mean-field error $\Delta E_\mathrm{MF} = E_\mathrm{MF}[\rho_\mathrm{DFA}] - E_\mathrm{MF}[\rho_\mathrm{HBCI}]$,
in kcal/mol, of self-consistent densities, with $E_\mathrm{MF} = T_s + E_\mathrm{ext} + E_\mathrm{H}$.}
\label{tab:mean_field_error}

\begin{tabular*}{\textwidth}{@{\extracolsep{\fill}}lrrrrrr@{}}
\toprule
System & PBE & SCAN & r2SCAN & B3LYP & NNGGA-UEG & NNLap \\
\midrule
O                & $-11.07$ & $1.67$  & $0.33$  & $-4.87$ & $-9.57$  & $-10.83$ \\
F                & $-12.69$ & $2.49$  & $0.76$  & $-5.60$ & $-11.39$ & $-10.80$ \\
H$_2$ (2eq)\footnote{H$_2$ at twice the equilibrium bond length ($r = 1.5$~\AA).}
                 & $-4.79$  & $-5.18$ & $-5.18$ & $-5.96$ & $-4.01$  & $-4.23$ \\
BH               & $3.77$   & $8.71$  & $8.58$  & $7.14$  & $5.84$   & $4.54$ \\
H$_2$O           & $-6.32$  & $3.39$  & $2.27$  & $-1.42$ & $-4.74$  & $-5.04$ \\
CH$_2$ (triplet) & $-7.58$  & $1.63$  & $1.06$  & $-4.53$ & $-5.30$  & $-7.20$ \\
CH$_2$ (singlet) & $0.54$   & $7.07$  & $6.64$  & $4.19$  & $2.69$   & $1.64$ \\
CN& $-10.23$ & $-0.23$ & $-1.22$ & $-8.52$ & $-7.12$  & $-7.72$ \\
\midrule
MAE              & $7.12$   & $3.80$  & $3.26$  & $5.28$  & $6.33$   & $6.50$ \\
\bottomrule
\end{tabular*}
\end{table*}

To assess whether the trends observed for W4-11 and BH76 extend to a
broader range of thermochemical problems, we consider 19 of the 55
component benchmark sets in GMTKN55~\cite{goerigk2017gmtkn55}.
Following Kanungo \textit{et al.}~\cite{kanungo2025learning}, we select the sets for which dispersion effects are negligible: those whose MAE with B3LYP changes by less than 30\% upon adding the D3 dispersion correction~\cite{grimme2010d3,grimme2011damping}.
These 19 dispersion-free sets belong to two GMTKN55 categories: basic properties and reaction energies of small systems (SMALL) and reaction barrier heights (BARRIER).
Because only 19 of the original 55 GMTKN55 sets
are included, the normalization factors entering the weighted total mean absolute deviation (WTMAD) metrics are
redefined for the selected sets.

The modified definitions are given in Appendix~\ref{app:wtmad}.
In brief, WTMAD-1 applies a discrete weight based on the characteristic energy scale of each component set, whereas WTMAD-2 additionally accounts for the number of reactions and rescales each set by its mean absolute reference energy.

Figure~\ref{fig:wtmad_two_panel} reports the modified WTMAD-1 and WTMAD-2 errors for the SMALL and BARRIER groups separately, as well as for their combined SMALL+BARRIER group.
For both metrics, NNGGA-UEG consistently improves upon PBE, while NNLap provides a further reduction in the error for the SMALL, BARRIER, and combined SMALL+BARRIER groups. The Laplacian-level model is competitive with SCAN and r\(^{2}\)SCAN and, in several cases,
yields lower errors, most notably for the BARRIER group under the WTMAD-1 measure. The hybrid functionals nevertheless remain substantially more accurate, with their largest advantage occurring for the barrier-height sets. This behavior is consistent with the
importance of nonlocal exchange for transition states involving stretched and weak bonds. Overall, the combined SMALL+BARRIER results show that the reduced-Laplacian dependence improves the balance between
small-molecule thermochemistry and reaction barriers, rather than producing an improvement confined to the W4-11 atomization energies.

\subsection{Total energies}

The thermochemistry tests above probe only energy differences.
Therefore, we also measure total energies, which provide a stricter test for our functional.
They probe whether the accuracy achieved stems from a serendipitous cancellation of errors in energy differences or from a better prediction of individual energies.
Since exact total energies from CI are available only for light atoms and a few small molecules, following Kanungo \textit{et al.}~\cite{kanungo2025learning}, we calculate the reference total energy of a molecule as $E_{\mathrm{ref}} = \sum_{A} n_A E_A - \mathrm{AE}$, where $E_A$ is the CI energy of atom type $A$ obtained from basis-set extrapolation, $n_A$ is the number of atoms of that type, and $\mathrm{AE}$ is the atomization energy from the W4-11 dataset~\cite{karton2011w4}.
This construction restricts the test to the 99 W4-11 molecules whose constituent atoms belong to the first two rows of the periodic table.

Figure~\ref{fig:total_energy} shows the errors in the total energy per atom for individual molecules and the corresponding MAEs.
NNLap reaches an MAE of 2.3 kcal/mol per atom, down from 17.6 kcal/mol for PBE and 5.8 kcal/mol for the gradient-level NNGGA-UEG.
It is comparable to SCAN (2.5 kcal/mol) and lower than r2SCAN (3.9 kcal/mol) and the hybrid B3LYP (8.9 kcal/mol).
These results indicate that a Laplacian-level NN model, trained on only a few systems, improves on the atomization energy benchmark due to more accurate energy predictions for the individual systems, and not just systematic error cancellation.

\subsection{Quality of density}

In addition to testing the total energies, we also check the self-consistent densities, using the mean-field error~\cite{gould2023step,kanungo2025learning}. For a density $\rho$, the mean-field energy $E_{\rm MF}[\rho]=\sum_\sigma T_s[\rho^\sigma]+E_{\rm H}[\rho]+\int\rho(\mathbf r)\,v_{\rm ext}(\mathbf r)\,d\mathbf r$ is the non-XC part of the total energy.
The mean-field error of a self-consistent density $\rho_{\rm DFA}$ is $\Delta E_{\rm MF}=E_{\rm MF}[\rho_{\rm DFA}]-E_{\rm MF}[\rho_{\rm ref}]$, with $\rho_{\rm ref}$ the CI density.
A common alternative for assessing the density is the density-driven error~\cite{kim2013understanding}.
It depends on the density as well as on the XC functional, so it does not isolate the error in the density itself.
So, we use the mean-field error which depends on the density alone and weights the density error by its effect on the energy.
Since $T_s[\rho_{\rm ref}^\sigma]$ requires the inverse DFT orbitals, we use the same eight systems as Kanungo \textit{et al.}~\cite{kanungo2025learning}, of which only O and triplet CH$_2$ belong to the training set.
Table~\ref{tab:mean_field_error} lists the results, with the computational details given in Appendix~\ref{app:scf}.

NNLap improves on PBE for six of the eight systems and lowers the MAE from 7.12 to 6.50 kcal/mol, on par with the 6.33 kcal/mol of NNGGA-UEG.
The Laplacian term therefore improves the total energies of Fig.~\ref{fig:total_energy} while retaining the density quality of the gradient-level model.
The meta-GGAs and B3LYP remain more accurate by this measure, with MAEs of 3.80, 3.26, and 5.28 kcal/mol.
For stretched H$_2$, all six functionals give nearly the same error.
The gain of both NN models over PBE comes from using the potential loss of Eq.~\eqref{eq:nn-potential-loss} in training, which drives the self-consistent solution to a better density than PBE.
It also generalizes beyond the training set: F, H$_2$O, CN, and stretched H$_2$ all improve over PBE.

\section{\label{sec:Conclusion} Conclusion}

We have presented NNLap, a Laplacian-level XC functional constructed as an NN correction to PBE and trained on exact XC potentials and energies of just five atoms and three molecules, all obtained from inverse DFT calculations.  
Adding the reduced Laplacian to the gradient-level model brings the largest gains for total and atomization energies.
The MAE of the total energy per atom decreases from 5.8 kcal/mol for NNGGA-UEG to 2.3 kcal/mol, better than higher rungs models such as SCAN, r2SCAN and B3LYP.
On W4-11, the MAE decreases from 9.357 to 6.477 kcal/mol, less than half the PBE value.
On the 19 dispersion-free subsets of GMTKN55, NNLap has similar accuracy as SCAN and r2SCAN.
These results show that a semilocal functional trained on a handful of systems can reach meta-GGA accuracy for energies through an explicit dependence on the density alone.
The Laplacian avoids the orbital dependence of $\tau$-based meta-GGAs and costs far less than exact exchange.

Although NNLap improves on PBE for every benchmark considered here, reaction barriers remain the least improved.
On BH76, the MAE decreases only from 9.151 kcal/mol for PBE to 8.001 kcal/mol, and the hybrid functionals remain substantially more accurate.
Transition states with stretched bonds appear to require nonlocal exchange, which semilocal ingredients like gradient and Laplacian of the density cannot provide.
In addition, the training set contains only atoms and small molecules at equilibrium, so the transfer of the functional to solids remains untested.
We intend to address both limitations in future work, by extending the training data and by building on the present framework toward functionals with nonlocal ingredients.

\begin{acknowledgments}
We acknowledge the support of Department of Energy, Office of Science, through grant number DE-SC0022241, under the auspices of which the inverse DFT calculations and the machine learning was conducted. We also acknowledge the support from AFOSR grant FA9550-24-1-0344 that supported the integration of the inverse DFT, the machine-learning, and the PySCF based testing framework. This study used resources of the NERSC Center, a DOE Office of Science User Facility supported by the Office of Science of the U.S. Department of Energy under Contract No. DE-AC02-05CH11231.\end{acknowledgments}

\appendix

\section{\label{app:scf} SCF calculations}

We perform self-consistent Kohn--Sham calculations with the trained NNLap functional in PySCF, using a custom implementation of the NN-based XC energy and potential.
The partial derivatives of the NN correction to $e_{\rm xc}$ with respect to $\rho^\sigma$, $\nabla\rho$, and $\nabla^2\rho$ are obtained with PyTorch automatic differentiation.
The native PySCF implementation handles the first two terms of Eq.~\eqref{eq:nn-vxc-explicit}.
The Laplacian term is not available in PySCF, so we implement our own finite-difference-based evaluation of this term and integrate it into the SCF cycle.
The thermochemistry benchmarking of the GMTKN55 sets uses the def2-QZVP basis set and the PySCF numerical integration grid at level 3.
The total energies of Fig.~\ref{fig:total_energy} and the density checks of Table~\ref{tab:mean_field_error} use the more refined cc-pV6Z basis set and grid level 9.
All calculations use an SCF energy convergence threshold of $10^{-8}$ Hartree.

\section{\label{app:wtmad} WTMAD metrics}

This appendix defines the modified WTMAD-1 and WTMAD-2 metrics used in Sec.~\ref{sec:Results} for the 19 dispersion-free GMTKN55 sets.
For a benchmark group \(\mathcal{G}\), the mean absolute error of its
\(i\)th component set is
\begin{equation}
\mathrm{MAE}_i
=
\frac{1}{N_i}
\sum_{j=1}^{N_i}
\left|
\Delta E_{ij}^{\mathrm{calc}}
-
\Delta E_{ij}^{\mathrm{ref}}
\right|,
\end{equation}
where \(N_i\) is the number of reactions in set \(i\). The
characteristic energy scale of that set is defined as
\begin{equation}
\overline{|\Delta E|}_i
=
\frac{1}{N_i}
\sum_{j=1}^{N_i}
\left|
\Delta E_{ij}^{\mathrm{ref}}
\right|.
\end{equation}
The modified WTMAD-1 metric is
\begin{equation}
\mathrm{WTMAD\text{-}1}_{\mathcal{G}}
=
\frac{1}{N_{\mathrm{set}}^{\mathcal{G}}}
\sum_{i\in\mathcal{G}}
w_i\,\mathrm{MAE}_i,
\end{equation}
where \(N_{\mathrm{set}}^{\mathcal{G}}\) is the number of component
sets in group \(\mathcal{G}\), and
\begin{equation}
w_i =
\begin{cases}
10,  & \overline{|\Delta E|}_i < 7.5~\mathrm{kcal/mol},\\
1,   & 7.5~\mathrm{kcal/mol}
       \leq \overline{|\Delta E|}_i
       \leq 75~\mathrm{kcal/mol},\\
0.1, & \overline{|\Delta E|}_i > 75~\mathrm{kcal/mol}.
\end{cases}
\end{equation}
The modified WTMAD-2 metric is
\begin{equation}
\begin{aligned}
\mathrm{WTMAD\text{-}2}_{\mathcal{G}}
&=
\frac{\overline{|\Delta E|}_{\mathcal{G}}}
     {N_{\mathrm{tot}}^{\mathcal{G}}}
\sum_{i\in\mathcal{G}}
N_i
\frac{\mathrm{MAE}_i}
     {\overline{|\Delta E|}_i},
\\
N_{\mathrm{tot}}^{\mathcal{G}}
&=
\sum_{i\in\mathcal{G}} N_i,
\\
\overline{|\Delta E|}_{\mathcal{G}}
&=
\frac{1}{N_{\mathrm{set}}^{\mathcal{G}}}
\sum_{i\in\mathcal{G}}
\overline{|\Delta E|}_i .
\end{aligned}
\end{equation}
For the combined SMALL+BARRIER group,
\(N_{\mathrm{set}}^{\mathcal{G}}=19\), while the SMALL and BARRIER
values are evaluated using only the component sets belonging to the
respective category.


\bibliography{MainPaperRefs}

\end{document}